\documentclass{ws-ijmpa}
\usepackage{amsmath}
\usepackage{amssymb}
\usepackage{graphicx}

\begin{document}

\markboth{Kevin Stenson}
{Hadronic charm mixing results from FOCUS}

%
\catchline{}{}{}{}{}
%

\title{$D^0$--$\overline{D}{}^0$ hadronic mixing and DCS decays from FOCUS}
\author{\footnotesize Kevin Stenson\footnote{on behalf of the FOCUS Collaboration (http://www-focus.fnal.gov/)}}
\address{Department of Physics, University of Colorado, Campus Box 390\\
Boulder, CO 80309, USA}

\maketitle

\begin{abstract}
We present an analysis of the decay $D^0\!\rightarrow\!K^+\pi^-$ from FOCUS\@.
From a sample of $234$ events we find a branching ratio of 
$\frac{\Gamma(D^0\!\rightarrow\!K^+\pi^-)}{\Gamma(D^0\!\rightarrow\!K^-\pi^+)} = 
(0.430^{\,+\,0.062}_{\,-\,0.061}\pm 0.031)\%$ under the assumption of no mixing and no \textit{CP} 
violation.  We also present limits on charm mixing.
\end{abstract}

\section{Introduction and event selection}

Charm mixing remains elusive.  In the Standard Model the charm mixing rate
is greatly suppressed by small CKM matrix
elements and strong GIM suppression.  This very fact, however, provides a unique opportunity
to search for new physics.

The search for charm mixing requires tagging the flavor of a neutral charm meson at production and 
again tagging the flavor at decay.  The production flavor is determined using $D^{*+}\!\rightarrow\!D^0\pi_s^+$
decays. The decay flavor is determined by reconstructing a Cabibbo favored (CF) decay with a charged kaon such as 
$D^0\!\rightarrow\!K^-\pi^+$.  A right-sign (RS) decay is defined as one in which the kaon and soft pion charges are 
opposite while same kaon and $\pi_s$ charges are wrong-sign (WS) decays.  The $D^0$ can also decay directly to 
$K^+\pi^-$ via a doubly Cabibbo suppressed decay (DCSD).  The time dependent WS decay rate can be 
written as:\\[-12pt]
\begin{equation}
\vspace{-1pt}
\label{eq:wsrate}
R_{WS}(t) = e^{-\Gamma t}\left( R_D + \sqrt{R_D} y' \Gamma t + \frac{1}{4}\left({x'}^2+{y'}^2\right) \Gamma^2 t^2\right)
\end{equation}
where the three terms correspond to DCSD, interference between mixing and DCSD, and mixing.
$R_D$ is the DCS branching ratio relative to the Cabibbo favored mode.  The paramters
$x' \equiv x \cos{\delta_{K\pi}} + y \sin{\delta_{K\pi}}$ and $y' \equiv y \cos{\delta_{K\pi}} - x \sin{\delta_{K\pi}}$ 
are rotated versions of the mixing parameters $x \equiv \frac{\Delta M}{\Gamma}$ and 
$y \equiv \frac{\Delta \Gamma}{2\Gamma}$ (mass and lifetime
splitting terms) and $\delta_{K\pi}$ is the strong phase between CF and DCS decay.
Discovery of hadronic charm mixing requires separating the three components 
of Eq.~\ref{eq:wsrate} using the different lifetime distributions.  Note that Eq.~\ref{eq:wsrate} has no information
on the sign of $x'$ and therefore the relevent fit 
variable is ${x'}^2$.  In this analysis, \textit{CP} violation effects are not considered and charge conjugates are implied.

FOCUS recorded data during the 1996--7 fixed-target run at Fermilab.  
A photon beam impinged on BeO targets.  
16 silicon strip planes provide vertexing and tracking.
Charged particles are tracked and momentum analyzed as they pass
through up to two dipole magnets and up to five sets of multiwire proportional chambers.  
Three \v{C}erenkov counters, two EM calorimeters, and two muon detectors identify particles.
A candidate driven vertexing algorithm is used to reconstruct charm.  For
$D^0\!\rightarrow\!K^-\pi^+$, two tracks must verticize with
CL $>$ 2\%.  This candidate is used to seed the 
production vertex which must have CL $>$ 1\%.  The soft pion $\pi_s^+$ from the 
$D^{*+}\!\rightarrow\!D^0\pi_s^+$ must
be consistent with originating from the production vertex and the track is refit to the production
vertex.  
Separating charm from hadronic background is primarily accomplished by requiring the decay
vertex be distinct from the production vertex.  Other cuts are made to
ensure the vertex is isolated and is more charm like than background like.
Strong \v{C}erenkov cuts are applied to the combination of $K^-$ and $\pi^+$ especially if the
invariant mass under a reflection of $K^-\!\rightarrow\!\pi^-$ and $\pi^+\!\rightarrow\!K^+$ is
close to the $D^0$ mass.
If more than one $D^{*+}$ candidate is found for one $D^0$ candidate an additional
cut may be made.
If a RS (WS) candidate is found with $2.5\,\textrm{MeV} < Q(D^*) < 9.5\,\textrm{MeV}$ 
then no WS (RS) candidates are allowed.  

\section{Fit description and results}

A 3D binned maximum likelihood fit is used in this analysis.  Two dimensions, 
$M(D^0)$ and $Q(D^*)$ separate signal from background while
$\tau(D^0)$ distringuishes DCSD, mixing, and intereference.  The RS
and WS data are fit simultaneously.
The fit components generally have a shape determined by a 
Monte Carlo simulation and a yield which is free to float, some with weak constraints imposed.  The 
components making up the fit which are modeled by the Monte Carlo are RS signal, 
WS signal (DCSD, interference, and mixing), real 
$D^0\!\rightarrow\!K^-\pi^+$ decay with a fake soft pion.
reflections to both RS and WS ($K^-K^+$, $\pi^-\pi^+$, $\pi^-\pi^+\pi^0$, $K^0\pi^-\pi^+$), RS background
($K^-\ell^+\nu$ and $K^-\pi^+\pi^0$), and WS background (double misidentification of $K^-\ell^+\nu$, 
$K^-\pi^+\pi^0$, $K^-\pi^+$).  The remaining background is some combination of non-charm and poorly
reconstructed charm
events.  This is modeled with functional forms in mass 
$a \exp{\left(b m\right)}$, energy release $\alpha q^{1/2} + \beta q^{3/2}$, and lifetime 
$\exp{\left(-t/\tau_1\right)} + \eta \exp{\left(-t/\tau_2\right)}$.  
Penalty terms are added to ensure backgrounds
are consistent with known branching ratios.  The affected backgrounds are the $D^0$ decays
to $K^+K^-$, $\pi^+\pi^-$, $\pi^+\pi^-\pi^0$, $K^0\pi^+\pi^-$, $K^-\pi^+\pi^0$, and $K^-\ell^+\nu$ which
all appear in RS and WS\@. 
The background due to double misid of the RS signal
is fixed based on the RS yield and relative efficiency.


Mini Monte Carlo tests of the fit were used to verify the accuracy of the reported fit errors and to
check for possible biases in the fit.  No significant bias was found and the reported fit errors agree
with the mini Monte Carlo results.
Fit variants with different binning, different constraints, and different accounting of the random
background were tried.  No significant differences were observed.  Variations of all the selection
criteria were also analyzed with no significant differences in the results.  For the branching ratios
reported, systematic errors due to fit variants and cut variants were obtained 
from the r.m.s of the variations and then added in quadrature.  

\begin{figure}
\centerline{\includegraphics[width=2.48in,height=1.93in]{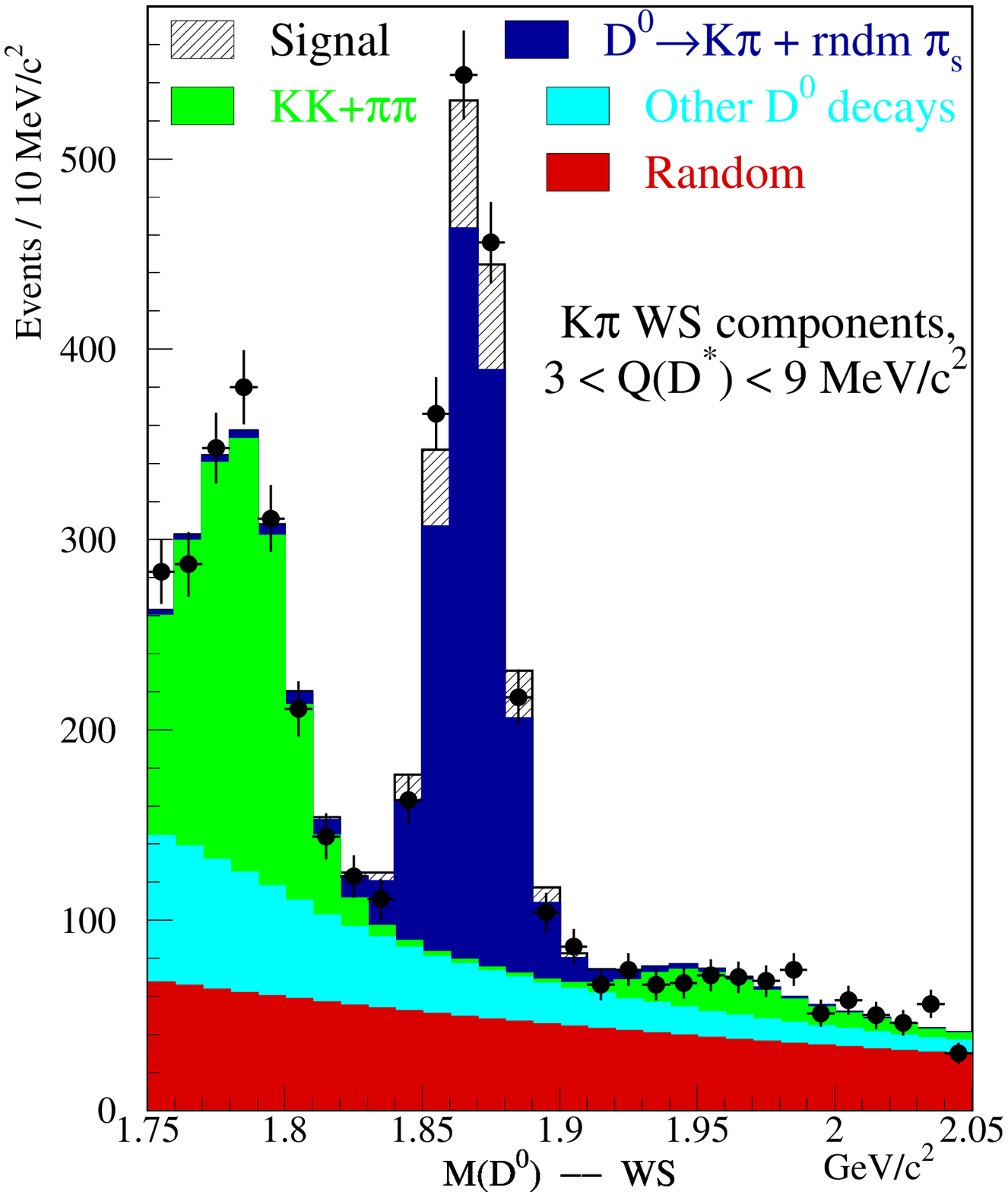}\hspace{-5pt}
\includegraphics[width=2.48in,height=1.93in]{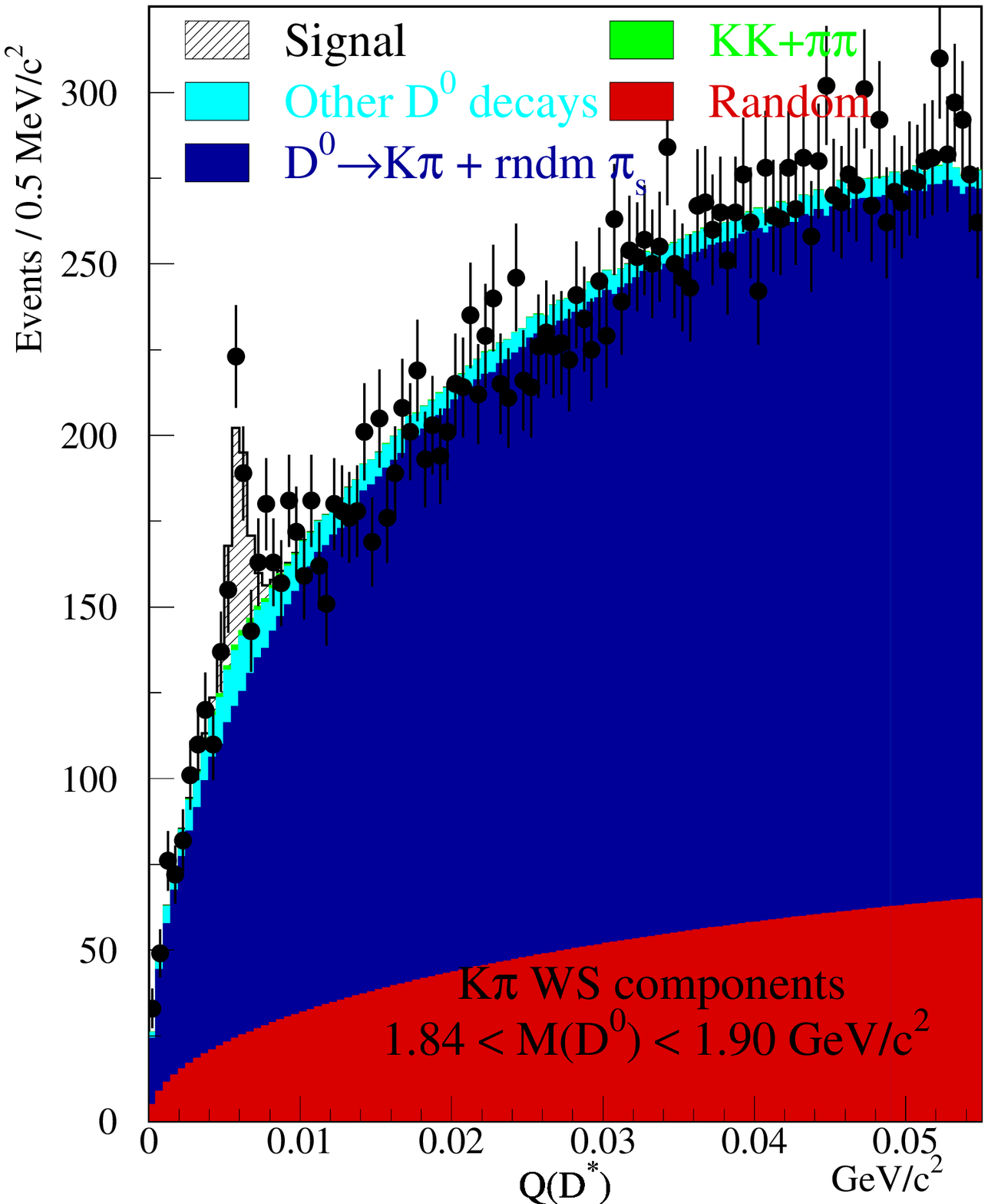}}
\centerline{\includegraphics[width=2.48in,height=1.75in]{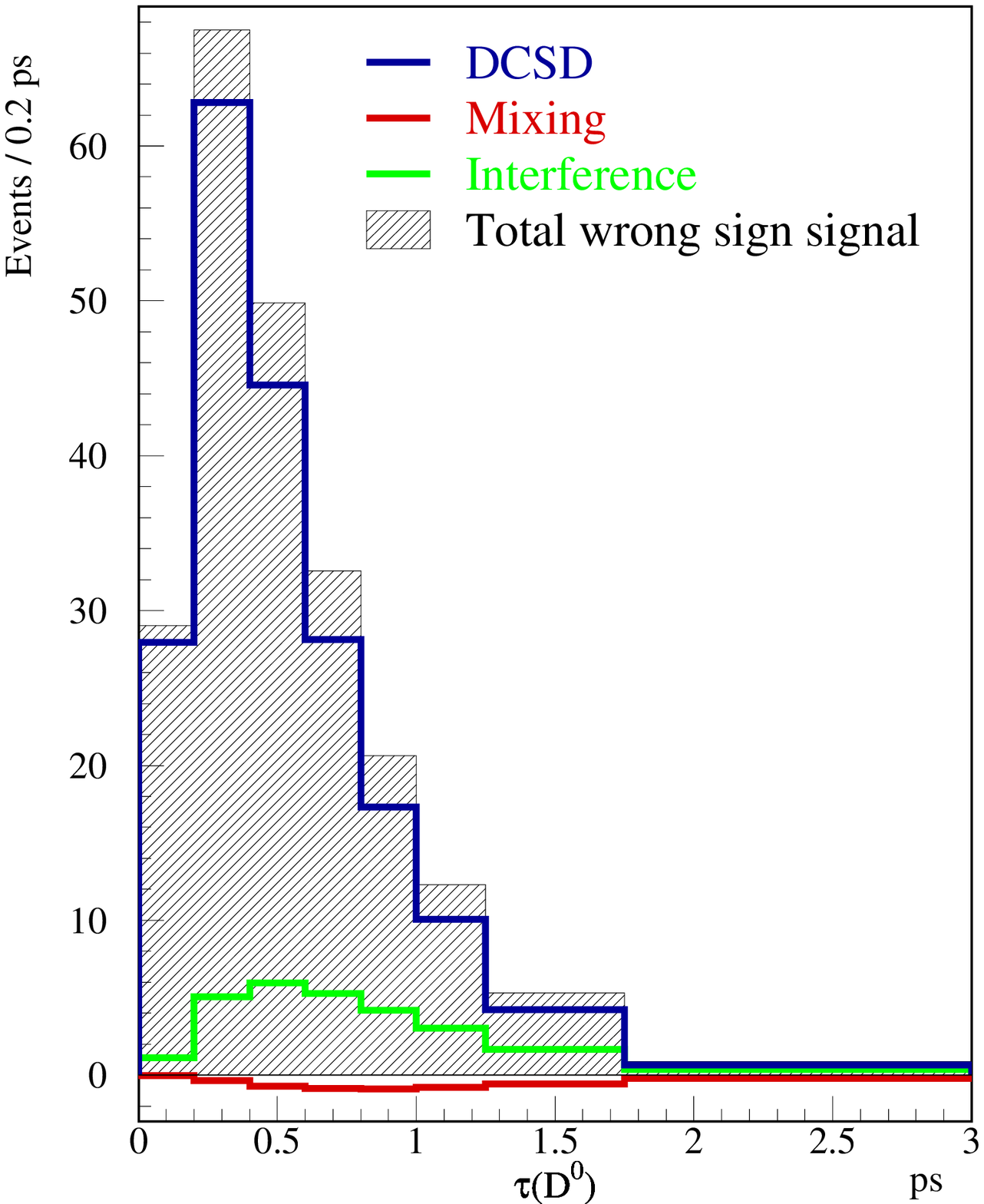}\hspace{-5pt}
\includegraphics[width=2.48in,height=1.75in]{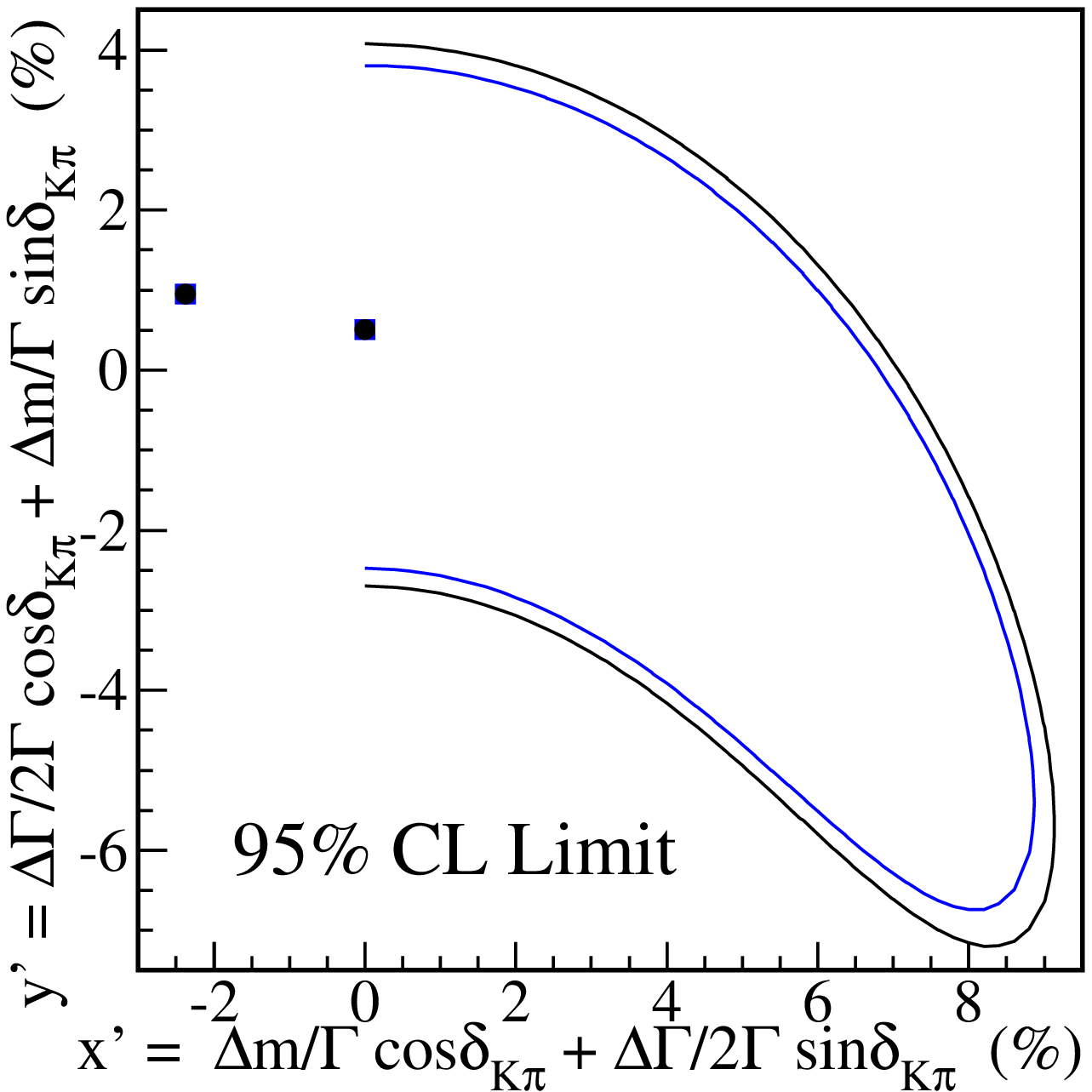}}
\caption{The top plots show projections onto $M(D^0)$ and $Q(D^*)$ of the fit and data for WS events.
The bottom-left plot shows the contributions to the WS signal versus $\tau(D^0)$. 
The bottom-right plots shows the 95\% CL ${x'},{y'}$ contour with statistical (inner) and full
errors (outer).}
\label{fig:mqt}
\end{figure}

Fitting without mixing we find
$R_{WS} = (0.430^{\,+\,0.062}_{\,-\,0.061}\pm 0.031)\%$. 
From the standard mixing fit we find a DCS branching ratio of 
$R_D = (0.382^{\,+\,0.167}_{\,-\,0.163}\pm0.069)\%$.
Figure~\ref{fig:mqt} shows the WS projections onto $M(D^0)$ and $Q(D^*)$.  
The RS data is dominated by $54452 \pm 242$ signal events.
Due to significant correlations between 
${x'}$ and $y'$, the interesting physics result is
a contour in the ${x'}$, $y'$ plane.  Although we fit for ${x'}^2$, we choose to plot ${x'}$.
The 95\% CL contour is defined as the location in the 
$x'$, $y'$ plane where $\Delta \log{\mathcal{L}} = 2.996$ relative to the minimum 
$-\log{\mathcal{L}}$ in the physical region of the $x'$, $y'$ plane.  
While the true minimum occurs at ${x'}^2 = -0.06\%$ and $y' = 1.0\%$,  
the minimum with ${x'}^2 \ge 0$ occurs at ${x'}^2 = 0$ and $y' = 0.5\%$ 
with a change in $-\log{\mathcal{L}}$ of only $0.006$ relative to the true minimum.  Systematic checks were
performed with 120 fit and cut variants.  The contour variation is consistent with differences in the
returned value of ${x'}^2$ and $y'$.  For the systematic error, we first find the change 
in $-\log{\mathcal{L}}$ between the global minimum and the ${x'}$, $y'$ location for each of the
variants.  We then find the value greater than 95\% of these differences which is $0.482$.  We
find the contour at which the change in $-\log{\mathcal{L}}$ is $2.996+0.482 = 3.478$ and
call this the 95\% CL including systematic error.  The $\tau(D^0)$ projection and $x',y'$ contours 
are shown in Fig.~\ref{fig:mqt}.  
Defining 95\% CL limits on $x'$ and $y'$ based on the projection of the
contour onto the respective axis we find ${x'}^2 < 0.83\%$ and $7.2\% < {y'} < 4.1\%$.


\end{document}